\documentstyle[prl,aps,psfig,floats,amsfonts,amssymb]{revtex}
\begin{document}
 \twocolumn[\hsize\textwidth\columnwidth\hsize\csname  @twocolumnfalse\endcsname
%
\draft
\title{Kondo effect in a magnetic field and the
magnetoresistivity of Kondo alloys}
\author
{T. A. Costi}
\address
{Institut Laue--Langevin, 6 rue Jules Horowitz, B.P. 156, 38042 Grenoble Cedex
9, France
}
\maketitle
\begin{abstract}
The effect of a magnetic field on the spectral density of a  $\rm{S=1/2}$
Kondo impurity is investigated at zero and finite temperatures by using
Wilson's numerical renormalization group method. A splitting of the total
spectral density is found for fields larger than
a critical value $H_{c}(T=0)\approx 0.5 T_{K}$, where $T_{K}$ is the 
Kondo scale. The splitting 
correlates with a peak in the magnetoresistivity of dilute magnetic 
alloys which we calculate and compare with the experiments on 
$\rm{Ce_{x}La_{1-x}Al_{2}}, x=0.0063$. 
The linear magnetoconductance of quantum dots 
exhibiting the Kondo effect is also calculated.
\end{abstract}
\vskip2pc]


The Kondo effect of a magnetic impurity interacting 
antiferromagnetically with conduction electrons has been the 
subject of theoretical and experimental 
investigations for several decades\cite{hewson.93}. 
It is most clearly manifested in the 
resistivity of electrons scattering from magnetic impurities in dilute
magnetic alloys, such as $\rm{Ce_{x}La_{1-x}Al_{2}}, x=0.0063$
\cite{felsch+winzer.73}, in which the resistivity
shows an anomalous increase with decreasing temperature below a very low 
temperature $T_{K}$, the Kondo temperature, which is a scale determined by
the many--body interactions. The phenomenon has recently acquired renewed
interest in several contexts including transport through strongly interacting 
quantum dots\cite{goldhaber.98}, and in the 
context of the dynamical mean field theory of strongly correlated 
lattice models\cite{georges.96}. Although
the thermodynamics of the Kondo model is well understood from exact 
solutions via the Bethe--Ansatz (BA)\cite{ba-kondo.80}, spectral and transport 
properties have proven more difficult to calculate reliably, especially
in the crossover region $T\approx T_K$ where, neither perturbation theory
about the Fermi liquid fixed point at $T\ll T_K$, nor perturbation theory 
about the free fixed point at $T\gg T_K$, is adequate. A non--perturbative
approach, such as the numerical renormalization group (NRG) method
\cite{wilson.75+kww.80}, is required to describe this crossover region. 
Recent developments in the NRG have yielded the zero field spectral 
and transport properties of the Anderson model in the Kondo regime
\cite{costi.94,costi.99}. However, at finite magnetic fields, there 
are still very few calculations of these  properties for a Kondo 
impurity\cite{keiter.76,moore.00}.
In this paper we calculate the field and temperature dependence of 
the Kondo resonance and the magnetoresistivity of a $S=1/2$ Kondo impurity
directly from the Kondo model by using the NRG method. The results are 
compared with available experimental data on dilute magnetic alloys. 
We also calculate the equilibrium magnetoconductance of a quantum dot 
in the Kondo regime. 

{\em Model---}
Our starting point is the $S=1/2$ Kondo model, which, for the purposes of
this section, we write in the more general form of an anisotropic exchange
\cite{anderson.69}
\begin{eqnarray}
{\cal H} &=& \sum_{k,\sigma} \epsilon_{k}c_{k\sigma}^{\dagger}c_{k\sigma} + 
\frac{J_{\perp}}{2}\sum_{kk'}
        (c_{k\uparrow}^{\dagger}c_{k'\downarrow}S^{-} +
         c_{k\downarrow}^{\dagger}c_{k'\uparrow}S^{+})\nonumber\\
  &+& \frac{J_{\parallel}}{2}\sum_{kk'}
         (c_{k\uparrow}^{\dagger}c_{k'\uparrow} -
          c_{k\downarrow}^{\dagger}c_{k'\downarrow})S^{z} 
+ g\mu_{B}HS_{z},\label{eq:AKM}
\end{eqnarray}
The first term represents non--interacting conduction electrons and the
second and third terms represent an exchange interaction between a localized
spin $1/2$ and the conduction electrons with strength 
$J_{\perp},J_{\parallel}$. The magnetic field, $H$, is taken to 
couple only to the impurity spin\cite{note.gfactor}. In the following we set
$g=\mu_{B}=k_B=1$.
The magnetoresistivity can be obtained from the ${\cal T}$--matrix
for electrons of spin $\sigma$ scattering from a Kondo impurity
\cite{keiter.76}. This is
defined by the identity\cite{hewson.93}
$
{\cal G}_{k,k',\sigma}(\omega) = 
\delta_{k,k'}{\cal G}_{k,k',\sigma}^{0}(\omega)
+{\cal G}_{k,k,\sigma}^{0}(\omega){\cal T}_{\sigma}(\omega) 
{\cal G}_{k',k',\sigma}^{0}(\omega),
$
where ${\cal G}_{k,k',\sigma}(\omega)=\langle\langle c_{k,\sigma}; 
c_{k',\sigma}^{+}\rangle\rangle$
is the retarded conduction 
electron Green function and ${\cal G}_{k,k',\sigma}^{0}(\omega)$ is 
the corresponding unperturbed Green function. The ${\cal T}$--matrix is
local due the local nature of the exchange 
interaction in (\ref{eq:AKM}).
The impurity spectral density, $A(\omega,T,H)$, giving the Kondo resonance 
is the sum of up spin and down spin impurity spectral densities,
$A(\omega,T,H)=\sum_{\sigma}A_{\sigma}(\omega,T,H)$, where
\begin{eqnarray}
&&A_{\sigma}(\omega,T,H)
=-\frac{1}{\pi}{\rm Im}\;{\cal T}_{\sigma}(\omega+i\delta,T,H).
\label{eq:total-sd}
\end{eqnarray}
The linear response magnetoresistivity, $\rho(T,H)$, due to a small 
concentration, $c\ll 1$, of Kondo impurities is given by
\begin{eqnarray}
\rho^{-1}(T,H) &=&\frac{ne^2}{2m}
\sum_{\sigma}\int_{-\infty}^{+\infty} d\omega\; \tau_{\sigma}(\omega,T,H)
\left(-\frac{\partial f}{\partial\omega}\right),\label{eq:magneto}
\end{eqnarray}
where $\tau_{\sigma}^{-1}(\omega,T,H) = cA_{\sigma}(\omega,T,H)$ is the
transport time of electrons of spin $\sigma$, $f$ is the Fermi function 
and $m,n$ and $e$ are the mass, concentration and charge, respectively, 
of the conduction electrons. An expression for the 
${\cal T}$--matrix is easily derived by considering the equation 
of motion of the Green function ${\cal G}_{k,k',\sigma}$. The 
result is
\begin{eqnarray}
{\cal T}_{\sigma}(\omega)  
&=&  \frac{J_{\parallel}}{2}\langle S_{z}\rangle + 
\langle\langle O_{\sigma};O_{\sigma}^{\dagger}\rangle\rangle,\nonumber\\
O_{\sigma} & =& 
\frac{J_{\perp}}{2}c_{0,-\sigma}S^{-\sigma}+ \sigma\frac{J_{\parallel}}{2}c_{0,\sigma}S_{z}\nonumber
\end{eqnarray}
where $c_{0,\sigma}=\sum_{k}c_{k,\sigma}$. This expression is also useful for
calculating the properties of the Kondo lattice model via the 
dynamical mean field theory.

Knowledge of the ${\cal T}$--matrix also allows
the magnetoconductance of a quantum dot in the Kondo regime 
to be calculated. The current, $I$,
through such a quantum dot attached to two leads with chemical potentials 
$\pm eV/2$ in the presence of a magnetic field $H$ is given by
$I(T,H,V) = \frac{e}{\hbar}\sum_{\sigma}\int_{-\infty}^{+\infty} 
A_{\sigma}^{{\rm neq}}(\omega,T,H,V)(f(\omega+eV/2)-f(\omega-eV/2))$,
where $A_{\sigma}^{{\rm neq}}(\omega,T,H,V)$ 
is the spectral density of the dot 
generalized to a non--equilibrium situation $V>0$\cite{herschfield.91}. 
A non--equilibrium approach is required to calculate 
$A_{\sigma}^{{\rm neq}}(\omega,T,H,V)$\cite{rg-neq-approaches}. 
In experiments on quantum dots it is, however, 
possible to access the linear regime
where $|e|V\ll  T_{K},H$\cite{goldhaber.98} 
and the relevant quantity is then the linear 
magnetoconductance $G(T,H)=G(T,H,V=0) = \lim_{V\rightarrow 0}{dI}/{dV}$,
\begin{eqnarray}
G(T,H) &=&\frac{e^2}{\hbar}
\sum_{\sigma}\int_{-\infty}^{+\infty} d\omega\; A_{\sigma}(\omega,T,H)
\left(-\frac{\partial f(\omega)}{\partial\omega}\right).\label{eq:conductance}
\end{eqnarray}
NRG calculations for $G(T,H=0)$ have been compared with 
experimental results on quantum dots in the Kondo regime and good 
agreement has been found \cite{goldhaber.98}. Here, we present
the field dependence of $G(T,H)$ in the strong correlation limit 
Kondo regime of a quantum dot.

{\em Method---}
The technique we use to calculate $A_{\sigma}(\omega,T,H)$ is
Wilson's NRG method \cite{wilson.75+kww.80} extended
to finite temperature dynamics\cite{costi.94}. This is a non--perturbative
method which gives the excitations and eigenstates 
of ${\cal H}$ at a decreasing set of energy scales 
$\omega_{N}=\Lambda^{-\frac{N-1}{2}}, (N=1,2,\dots, \Lambda > 1)$, 
logarithmically spaced
about the Fermi level $\epsilon_{F}=0$. It thereby allows a direct 
calculation of dynamic quantities, such as $A_{\sigma}(\omega,T,H)$,
at a corresponding set of frequencies, $\omega\sim \omega_{N}$, and 
temperatures, $k_{B}T=T_{N}\sim \omega_{N}$, via the Lehmann 
representation. For further details see \cite{costi.99}. Accuracy 
is discussed below.

\begin{figure}[t]
\centerline{\psfig{figure=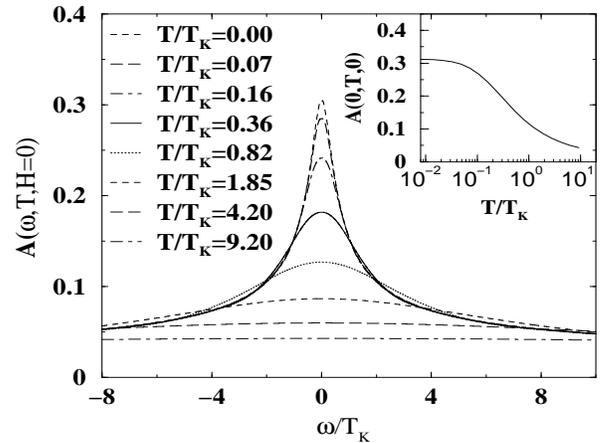,width=7.8cm,height=6.0cm,angle=0}}
\vspace{0.1cm}
\caption{
Temperature dependence of the Kondo resonance at $H=0$ calculated from 
Eq.\ (\ref{eq:total-sd}). The inset shows the peak height as a function of 
temperature.
}
\label{mr-fig1}
\end{figure}

{\em Calculations---}
We now restrict ourselves to the case of relevance to Kondo alloys 
and quantum dots, $J=J_{\perp}=J_{\parallel} \ll  D$, with $D=1$ 
the half--bandwidth of the unperturbed conduction electron density 
of states per spin, $N(\omega)$, which we 
take to be a semi--elliptic. The number of states retained per NRG iteration
was $462$ and a discretization parameter of $\Lambda=1.5$ was used. 
We used $J/D=0.3$, $N(0)J=0.19$ 
for all calculations, giving a Kondo scale of $T_{K}=0.0072$, as 
deduced from the half--width at half--maximum (HWHM) of the 
Kondo resonance at $T=0$ (Fig.\ \ref{mr-fig1}). We shall use this
definition of $T_K$ throughout this paper. As expected, it
is of the same order of magnitude as the expression 
$T_{K}'=D(N(0)J)^{1/2}e^{-1/N(0)J}\approx 0.0023$, which is valid for weak
coupling $N(0)J\ll  1$. 

{\em T--dependence of the Kondo resonance---}
The temperature dependence of the Kondo resonance in zero 
magnetic field is shown in Fig.\ \ref{mr-fig1}.
This is also known via NRG calculations on  
the Anderson impurity model\cite{costi.94}.  
By carrying out calculations on the Kondo model we can 
focus more explicitly on the many--body aspects of this temperature 
dependence. The inset shows that the height of the resonance,
which initially falls slowly at $T\ll T_{K}$ 
(as $A(0,0,0)(1-\alpha (T/T_K)^2$)), decreases approximately logarithmically
in a small range $0.1T_K<T<T_K$ close to $T_K$, with $T_{K}$ defined 
from the HWHM of the $T=0$ resonance. The resonance falls to half 
its $T=0$ peak height at $T\approx 0.5T_{K}$. The $\omega=T=0$ value 
satisfies, for both zero and finite magnetic fields (see 
Fig.\ \ref{mr-fig2} below), the Fermi liquid relation (Friedel sum rule) 
\begin{equation}
A_{\sigma}(\omega=0,T=0,H) = \frac{1}{\pi^{2} N(0)}\sin^{2}
\delta_{\sigma}(H),
\label{eq:FSR}
\end{equation}
where $\delta_{\sigma}(H)$ is the field dependent phase shift for 
electrons scattering from the impurity. This can be deduced
from the Anderson impurity model\cite{langreth.66}, 
of which the Kondo model is the low energy part, or, by assuming 
a local Fermi liquid description for the Kondo model\cite{nozieres.74}. 
It provides a useful test on the accuracy of the calculations.
On noting that the zero field phase shift in the Kondo model is $\pi/2$,
we have $A(0,0,0)=\sum_{\sigma}A_{\sigma}(0,0,0)=1/\pi=0.3183$. 
The numerical result gives $0.3109$ resulting in a relative error of 2\%. 

{\em H--dependence of the Kondo resonance---}The Kondo resonance ``splits''
in a large enough magnetic field---in the sense that the curvature in 
the {\em total} impurity spectral density at $\omega=0$ changes sign 
above a certain field, see Fig.\ \ref{mr-fig3}. 
Fig.\ \ref{mr-fig2} shows the evolution of the up spin part 
$A_{\uparrow}(\omega,T,H)$ at $T=0$ as a function of field
($A_{\downarrow}(\omega,T,H)=A_{\uparrow}(-\omega,T,H)$). 
The value at the Fermi level, given by 
Eq.\ (\ref{eq:FSR}), probes the field dependent 
phase shift $\delta_{\uparrow}(H)$. This has been calculated
exactly from the BA solution of the Kondo model
\cite{ba-kondo.80} by Andrei \cite{andrei.82}, via its relation,
$\delta_{\uparrow}(H)=\frac{\pi}{2}(1-2M_{i}(H))$,
to the impurity magnetization $M_{i}(H)$. The inset to
Fig.\ \ref{mr-fig2} shows very good agreement between
the NRG and BA for $A_{\uparrow}(\omega=0,T=0,H)$ for
fields up to $5T_{K}$. The same 2\% known error in the NRG result for the
Friedel sum rule at $H=0$ was assumed for all fields and corrected
for in the comparison. We estimate the shift, $E_0(H)$, of the maximum in 
$A_{\uparrow}(\omega,T=0,H)$ to be $E_0(H)=H$ at $H\ll T_K$ and 
$E_0(H)\approx 0.85H\pm0.1H$ for $H>T_K$. 
\begin{figure}[t]
\centerline{\psfig{figure=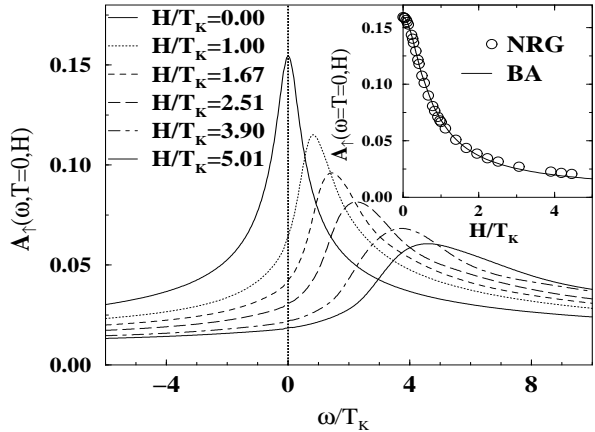,width=7.8cm,height=6.0cm,angle=0}}
\vspace{0.1cm}
\caption{
The up spin component of the Kondo resonance at $T=0$ for several magnetic
fields. The inset compares the NRG and BA (BA)
values for $A_{\uparrow}(\omega=T=0,H)$ for finite fields.
}
\label{mr-fig2}
\end{figure}
\begin{figure}[t]
\centerline{\psfig{figure=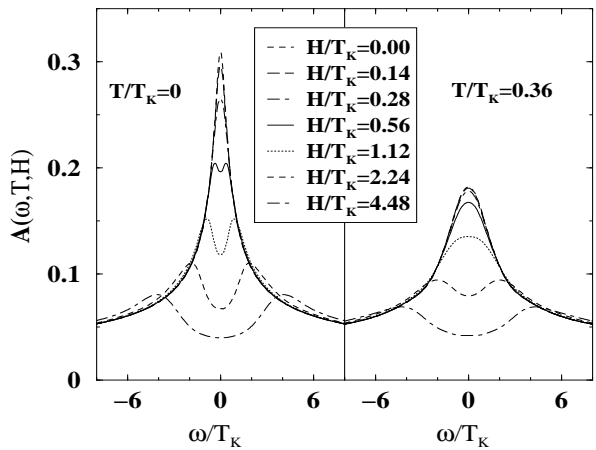,width=7.8cm,height=6.0cm,angle=0}}
\vspace{0.1cm}
\caption{
Splitting of the Kondo resonance with increasing magnetic field 
at $T=0$ and $T/T_{K}=0.36$. 
}
\label{mr-fig3}
\end{figure}

The Kondo resonance depends strongly on temperature and 
field, so it is not a priori clear that the total 
spectral density will split in any magnetic field either at $T=0$ or 
at higher temperatures. This depends on the competition
between the reduction in height of the up spin and down 
spin components 
and their outward shift as a function of field. That this 
happens above a critical field, $H_c$, can be seen from
Fig.\ \ref{mr-fig3} for two cases: $T=0$ and $T/T_{K}=0.36$. 
Experimental evidence for this splitting comes 
from the temperature
dependence of the magnetoresistivity of Kondo impurities, which we
shall discuss below. Point contact spectroscopy on dilute
Kondo alloys \cite{yanson.95} also indicates this splitting.
At very low temperatures $T < 0.25T_K$ we 
find  that $H_{c}$ is close to its $T=0$ value, $0.5T_K$, 
and at higher temperatures is linear in temperature 
($H_{c}(T)\approx 3T$). We can compare these results with
those obtained from a recent calculation of the one-spinon density of 
excitations for the Kondo model  at $T=0$ \cite{moore.00}. The 
low energy behaviour of this quantity should be close to the true spectral 
density $A_{\sigma}(\omega\rightarrow 0,T=0,H)$. For 
$H_{c}(T=0)$, which is a low energy property, the BA\cite{moore.00} gives
$g\mu_{B}H_{c}(T=0)=0.4858k_{B}T_{K}$ in excellent agreement with our 
value of $0.5k_{B}T_{K}$. Both approaches give a peak 
position $E_{0}(H)$ lying close to but always below $H$.

\begin{figure}[t]
\centerline{\psfig{figure=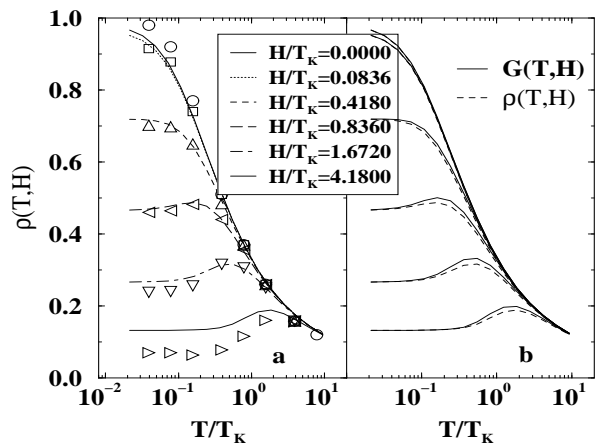,width=7.8cm,height=6.0cm,angle=0}}
\vspace{0.1cm}
\protect\caption{
(a) Comparison of the calculated magnetoresistivity curves (lines) and 
those of ${\rm Ce_{x}La_{1-x}Al_{2}, x=0.0063}$ (symbols)
{\protect\cite{felsch+winzer.73}}. 
Fields are in units of the theoretical $T_K$ and correspond to 
$H/kOe:0,1,5,10,20,50$ of {\protect\cite{felsch+winzer.73}} so our 
$T_K=0.8K$ ($T_K\approx 1K$ is cited in the experiment). (b) 
Comparison of the magnetoresistivity
and magnetoconductance through a Kondo impurity (quantum dot). 
All curves in (a) and (b) were scaled by their 
unitarity limits ($\rho(0,0)$ and $G(0,0)$).
}
\label{mr-fig4}
\end{figure}

{\em Magnetoresistivity---}
Fig.\ \ref{mr-fig4}a, shows the temperature dependence of the 
magnetoresistivity together with a comparison to experimental results on 
the dilute Kondo alloy ${\rm Ce_{x}La_{1-x}Al_{2}}, x=0.0063$
\cite{felsch+winzer.73}. We first describe the theoretical results.

The magnetoresistivity for $H=0$ shows the behaviour
known from calculations on the Anderson model\cite{costi.94},
including the $T^2$ Fermi liquid corrections to the resistivity from its 
unitarity limit at $T\ll T_K$\cite{nozieres.74} and the 
approximate logarithmic increase 
with decreasing temperature in a small range 
below $T\approx T_K$.  
On increasing the magnetic field,
a finite temperature peak appears at $T\approx H $ in
$\rho(T,H)$ for fields larger than $0.5T_K$. This reflects the 
splitting of the Kondo resonance, found above, for fields larger 
than $0.5T_K$ and is physically reasonable, since the resistivity 
from Eq.(\ref{eq:magneto}) can be thought of as a broadened measure
of the total spectral density. This statement can be made precise 
in the case of the conductance through a quantum dot in the Kondo regime, 
since from Eq.(\ref{eq:conductance}) we see that this quantity measures 
the field and temperature dependence of 
the Kondo resonance broadened by the derivative of the Fermi function. 
In Fig.\ \ref{mr-fig4}b we quantify the difference between the 
magnetoresistivity of a Kondo impurity and the magnetoconductance 
through such an impurity by comparing them. 
From Eq.(\ref{eq:magneto}--\ref{eq:conductance}) we see that they are equal at
$T=0$ and all fields. They differ slightly
only for fields comparable to or larger than $T_K$, and then only in the
temperature range around $T\approx H$. Hence, both quantities can 
be thought of as a ``spectroscopic'' measurement of the Kondo 
resonance\cite{herschfield.91}.

The dilute magnetic alloy ${\rm Ce_{x}La_{1-x}Al_{2}}, 
x=0.0063$ \cite{felsch+winzer.73} was the best candidate 
we could find for a comparison
of the magnetoresistivity to experimental results. 
It has a low lying doublet which can be modelled by a
$S=1/2$ Kondo model and both the normal part of the magnetoresistivity and 
the part due to disorder could be subtracted out 
\cite{felsch+winzer.73}. The resulting 
magnetoresistivity at the lowest temperature was 
compared with $T=0$ BA\cite{andrei.82} results
in \cite{schlottmann.87} and found to 
be in good agreement, except at fields very small 
and very large relative to $T_K$. 
The finite temperature comparisons in Fig.\ \ref{mr-fig4}a 
also show good agreement with the experimental results, except, 
as in the $T=0$ comparisons, at fields very small and 
very large relative to $T_K$ (see below). 
There are no free parameters in these comparisons. The curves were scaled 
by the unitarity value $\rho(0,0)$ and the relation between scales 
in experiment and theory was fixed by comparing the 
theoretical $\rho(T=0,H)$ versus
$H$ curve to the corresponding experimental one as in the comparisons
to the $T=0$ BA calculations. This scaling factor for fields
applies to all energies and was used to rescale the experimental 
temperatures onto the theoretical ones in units of our $T_K$. 
The large field dependence observed in the experiments between
$H=0$ and $H=1{\rm kOe}$ cases (at $T\ll T_K$) could 
be due to the finite field extrapolation used to obtain 
the $H=0$ case (required because ${\rm Ce_{x}La_{1-x}Al_{2}}, x=0.0063$ 
is superconducting at $H=0$ \cite{felsch+winzer.73}). 
The experiments also give too strong a suppression of
the magnetoresistivity with increasing field at large fields as compared
with both our calculations and with the $T=0$ BA results
\cite{schlottmann.87}. The origin of this is unclear, but it 
may indicate that the real system is more complicated than an effective
$S=1/2$ Kondo model at these large fields.

In summary, we have studied the splitting of the Kondo resonance in a
magnetic field by calculating the ${\cal T}$--matrix 
of the $S=1/2$ Kondo model and we have shown that
the splitting (in the total spectral density) occurs above a critical 
field $H_c(T=0)\approx 0.5T_K$. This is manifested in  
a peak at finite temperature for $H>0.5T_K$ in the magnetoresistivity 
of Kondo alloys and in the magnetoconductance 
of a quantum dot in the Kondo regime. Both quantities 
give ``spectroscopic'' information on the Kondo resonance and we
showed explicitly that they are almost equal for most fields and temperatures.
Quantitatively good agreement with the magnetoresistivity of the
Kondo alloy ${\rm Ce_{x}La_{1-x}Al_{2}}$ was found with 
no free parameters. Our results should be of use in interpreting
experiments on the field dependence of the conductance of quantum dots in
the Kondo regime.

Useful discussions with R. Chitra, A. Jerez, N. Manini, P. Nozi\`eres, 
F. Pistolesi, L. Taillefer and T. Ziman are gratefully acknowledged.
We also thank K. Winzer for sending us enlarged figures of the 
experimental data.


\begin{references}
\bibitem{hewson.93} A. C. Hewson, {\em The Kondo Problem to Heavy
Fermions} (Cambridge University Press, Cambridge, England 1993)
\bibitem{felsch+winzer.73} W. Felsch and K. Winzer, Solid State Comm. 
{\bf 13}, 569 (1973).
\bibitem{goldhaber.98} D. Goldhaber--Gordon et al., Phys. Rev. Lett. {\bf 81},
5225 (1998).
\bibitem{georges.96} A. Georges et al., Rev. Mod. Phys. {\bf 68}, 13 (1996).
\bibitem{ba-kondo.80} N. Andrei, Phys. Rev. Lett. {\bf 45}, 397 (1980);
P. B. Wiegmann, Sov. Phys. JETP. Lett. {\bf 31}, 364 (1980).
\bibitem{wilson.75+kww.80} K. G. Wilson,  Rev. Mod. Phys.{\bf 47}, 773
(1975); H. R. Krishnamurthy, J. W. Wilkins and K. G. Wilson,  
Phys. Rev. B{\bf 21}, 1003 (1980).
\bibitem{costi.94} T. A. Costi, A. C. Hewson and V. Zlati\'{c}, 
J. Phys. Cond. Matt. {\bf 6}, 2519 (1994).
\bibitem{costi.99} T. A. Costi, in {\em Density Matrix 
Renormalization}, edited by I. Peschel, X. Wang, M. Kaulke and 
K. Hallberg (Springer, Berlin, Germany 1999).
\bibitem{keiter.76} H. Keiter, Z. Phys. B{\bf 23}, 37 (1976); H. Keiter
and J. Kurkij\"{a}rvi, Z. Phys. B{\bf 26}, 169 (1977).
\bibitem{moore.00} J. Moore and X-G. Wen, cond-mat/9911068, and private
communication.
\bibitem{anderson.69} P. W. Anderson and G. Yuval, Phys. Rev. Lett. 
{\bf 23}, 89 (1969);  Phys. Rev. B{\bf 1}, 4464 (1970). 
\bibitem{note.gfactor} Setting the $g$-factor, $g_{e}$, of the 
electrons to zero allows a direct comparison with the experimental 
results in \cite{felsch+winzer.73} in which the small normal electronic 
contribution to the magnetoresistivity was already subtracted out.
For impurity systems, a finite $g_e$, will not affect the transport
time or magnetoresistivity appreciably  unless $g_{e}\mu_{B}H$ is 
comparable to the energy scale for variation of the conduction electron 
density of states at the Fermi level (a condition usually satisfied).
\bibitem{herschfield.91} S. Herschfield, J. H. Davies and J. W. Wilkins,
Phys. Rev. Lett. {\bf 67}, 3720 (1991); Y. Meir, N. S. Wingreen and P. A. Lee,
Phys. Rev. Lett. {\bf 70}, 2601 (1993). We assumed equal and energy 
independent couplings, of strength $\Gamma$ , to the left and right leads 
and included these in $A$.
\bibitem{rg-neq-approaches} H. Schoeller and J. K\"{o}nig, Phys. Rev. 
Lett. {\bf 84}, 3686 (2000); T. A. Costi,  Phys. Rev. B{\bf 55}, 3003 (1997).
\bibitem{langreth.66} D. C. Langreth, Phys. Rev. {\bf 150}, 516 (1966).
\bibitem{nozieres.74} P. Nozi\`eres, J. Low Temp. Phys. {\bf 17}, 31 (1974).
\bibitem{andrei.82} N. Andrei, Phys. Lett. {\bf 87A}, 299 (1982).
\bibitem{yanson.95} I. K. Yanson et al., Phys. Rev. Lett. {\bf 74}, 302 (1995).
\bibitem{schlottmann.87} P. Schlottmann, Phys. Rev. B {\bf 35}, 5279 (1987).
\end{references}
\end{document}